\pdfoutput=1
\documentclass{article}
\usepackage{caption}
\usepackage{subcaption}
\usepackage{graphicx}
\usepackage{amsmath}
\usepackage{geometry}
\usepackage{authblk}
\usepackage{hyperref}
\usepackage[T1]{fontenc}
\usepackage[utf8]{inputenc}
\usepackage{float}
\usepackage{csquotes}
\usepackage[version=4]{mhchem}
\usepackage{tikz}
	
\usetikzlibrary{backgrounds,fit,decorations.pathreplacing}% TikZ libraries
\title{Statistical correlation between quantum entanglement and spin-orbit coupling in crossed beam molecular dynamics}
\author{Junxu Li, Manas Sajjan, Sumit Suresh Kale, and Sabre Kais\thanks{Email: kais@purdue.edu}}
\affil{Department of Chemistry, Department of Physics and Astronomy, and

Purdue Quantum Science and Engineering Institute

Purdue University, West Lafayette, IN 47907, United States}

\begin{document}
\maketitle
	
\begin{abstract}
Non-classical features like interference is already being harnessed to control the output of chemical reactions. However quantum entanglement which is an equally enigmatic many-body quantum correlation can also be used as a powerful resource yet have eluded explicit attention. In this report, we propose an experimental scheme under the crossed beam molecular dynamical setup, with the $\ce{F}+\ce{HD}$ reaction, aiming to study the possible influence of entanglement within reactant pairs on the angular features of the product distribution. The aforesaid reaction has garnered interest recently as an unusual horseshoe shape pattern in the product ($\ce{HF}$) distribution was observed, which has been attributed to the coupling of spin and orbital degrees of freedom. We propose an experimental scheme aiming to study the possible influence of entanglement on the necessity for the inclusion of such spin-orbit characteristics, under circumstances wherein the existence of entanglement and spin-orbit interaction is simultaneously detectable.
We further numerically simulate the attainable results highlighting specific patterns corresponding to various possibilities. Such studies if extended can provide unforeseen mechanistic insight in analogous reactions too from the lens of quantum information.

\end{abstract}

\section*{Introduction}
\label{introduction}
The past decades have seen tremendous progress in the field of quantum information and quantum computing. 
With the expeditious developments in the hardware and software fronts, the algorithms developed on current state-of-the-art Quantum computers could help us to overcome the research obstacles that are beyond the capacity of the best available supercomputers\cite{whaley2014quantum, preskill2021quantum}.
The veritable workhorse of such algorithms are fundamental quantum properties such as superposition, entanglement, coherence, and interference which has been aptly exploited in secure communication \cite{ursin2007entanglement,zhang2017quantum}, to develop better sensing and meteorological tools \cite{degen2017quantum,pirandola2018advances,giovannetti2006quantum}, and even to unravel the mysteries of complex chemical processes by precisely simulating large chemical systems \cite{karra2016prospects,cao2019quantum,king2021scaling}.

Entanglement is a distinctive quantum mechanical feature that represents a strong correlation in many-body systems unexplainable by classical physics.
It exists at the core of quantum information processing and may prove essential for quantum speed-up\cite{kais2007entanglement}.
Several experiments realized entanglement in a wide range of physical systems, including trapped ions, quantum dots \cite{bayer2001coupling}, superconducting qubits \cite{shankar2013autonomously}, photons \cite{wang2016experimental}, between an atom and a molecule \cite{huang2005entanglement,lin2020quantum} and in complex chemical and biological systems\cite{huang2006entanglement,oh2008entanglement,zhu2012multipartite,pauls2013quantum}.
The coherent control of chemical reactions has been a long withstanding challenge in chemical physics and several studies have already found new powerful techniques that employs quantum superposition and interference to resolve this issue \cite{blasing2018observation,jambrina2015quantum, kale2021constructive}.
However, entanglement as a resource have received scanty attention so far.
In this report, we serve to address this lacuna and focus on gaining a deep insight on the correlations between pre-existing entanglement in a chemical reaction and how it affects the geometrical distribution of product formation. 

We choose to elucidate the implications using the
$\ce{F}+\ce{H2}$ reaction as a test-bed.
Such a choice is motivated by the fact that in the domain of resonances in chemical reactions, the $\ce{F}+\ce{H2}$ system, together with its said isotopic partner has been used to benchmark many studies\cite{zare1973direct,tully1974collisions,herschbach1977molecular,herschbach1987closing,levine2009molecular}.
Theoretical calculations first predicted the existence of reactive resonances for the $\ce{F}+\ce{H2}$ reaction in 1973 using the collinear reaction model\cite{schatz1973large,wu1973quantum}. Though such short-lived reactive resonance in the transition state of chemical reactions has been long predicted based on quantum dynamics
simulations \cite{truhlar1984resonances},
the direct characterization of transition states has been in the past a grand
challenge in physical chemistry \cite{fernandez2002scattering,polanyi1995direct},
due to the experimental absence of the characteristic Lorentzian signatures in the integral cross section (ICS) \cite{skodje2000resonance}. 
However in 1984, the forward-scattering peak potentially attributable to the reactive resonances for $\ce{HF}(v'=3)$ was first observed by Lee and coworkers in a crossed–molecular beam experiment performed on the $\ce{F}+\ce{H2}$ reactions\cite{neumark1984experimental}. 
Even though the study was pioneering yet the assertions remained inconclusive due possible existence of other dynamical origins for the forward scattering peak\cite{aoiz1994classical,castillo1996quantum}. 
To locate the reactive resonance, techniques of transition-state spectroscopy were later employed probing the dynamics near the transition state via the negative ion (\ce{FH2-}) photodetachment study\cite{manolopoulos1993transition, russell1996observe}.

The controversies of this scattering peak were finally resolved at the beginning of $21$ st century with the isotopic analogous reaction $\ce{F}+\ce{HD}$\cite{skodje2000resonance}. In the next two decades enormous enthusiasm was evoked investigating both the $\ce{F}+\ce{H2}$ and $\ce{F}+\ce{HD}$ reactions using various advanced technologies, such as the negative-ion photodetachment spectroscopic method\cite{kim2015spectroscopic,yang2019enhanced}, and the highly sensitive H atom Rydberg tagging time-of-flight method\cite{qiu2006observation,xu2006global,ren2008probing}. 
Even though, in all the above studies the experimental signature could be accurately explained based on the adiabatic theory thereby obviating the need to include special characteristics like the spin-orbit coupling. The situation however changed recently when a peculiar horseshoe-shaped pattern was observed in the product rotational-state–resolved differential cross section (DCS) in $\ce{F}({}^2P_{3/2})+\ce{HD}$ reactions, which the authors attribute to full spin-orbit characteristics\cite{chen2021quantum}. The remarkably unusual dynamical pattern provides a window into studying how the spin-orbit interaction can effectively influence chemical reactions.
Intuitively, there arises a question whether the pre-existence of entanglement between the reacting partners would have any influence on the spin-orbit interaction in the $\ce{F}+\ce{HD}$ reaction.
For instance, what if the incident F atoms are entangled pairs between the ground state ${}^2P_{3/2}$ and the excited state ${}^2P_{1/2}$.
Though the pioneers have already considered the inclusion of entanglement in chemical reactions\cite{zeman2004coherent,gong2003entanglement}, the possible correlation it may share with spin-orbit coupling have not been explicated as it is an inherently challenging study.
As a simple example, if we prepare a number of entangled F atom pairs dividing into two groups, then estimate the entanglement witness function measuring one group, while studying the chemical reactions using the other one,
the result can hardly be convincing due to the possibility that entanglement might get broken in collision. Thus, one must ensure that the spin-orbit interaction and entanglement can be detected simultaneously.
In this paper, we will propose a scheme for the simultaneous detection of spin-orbit characteristics and entanglement in the $\ce{F} + \ce{HD}\longrightarrow \ce{HF}+\ce{D}$ reactions.

The organization of the paper is as follows. In section \ref{Experimental Design} we propose the experimental setup
and discuss the implementation thoroughly. In sub-section \ref{SOC}
we review the need for the inclusion of spin-orbit coupling as illustrated in Ref.\cite{chen2021quantum}. In sub-section \ref{Entanglement_det} we elucidate how correlations due entanglement between the initial pair of $\ce{F}$ may be experimentally revealed and quantified. In section \ref{Num_sim} we simulate and analyze the various outcomes of the proposed experiment. We conclude thereafter with possible future implications.

\section{Experimental Design}
\label{Experimental Design}
\begin{figure}[ht]
    \centering
    \includegraphics[width=0.85\textwidth]{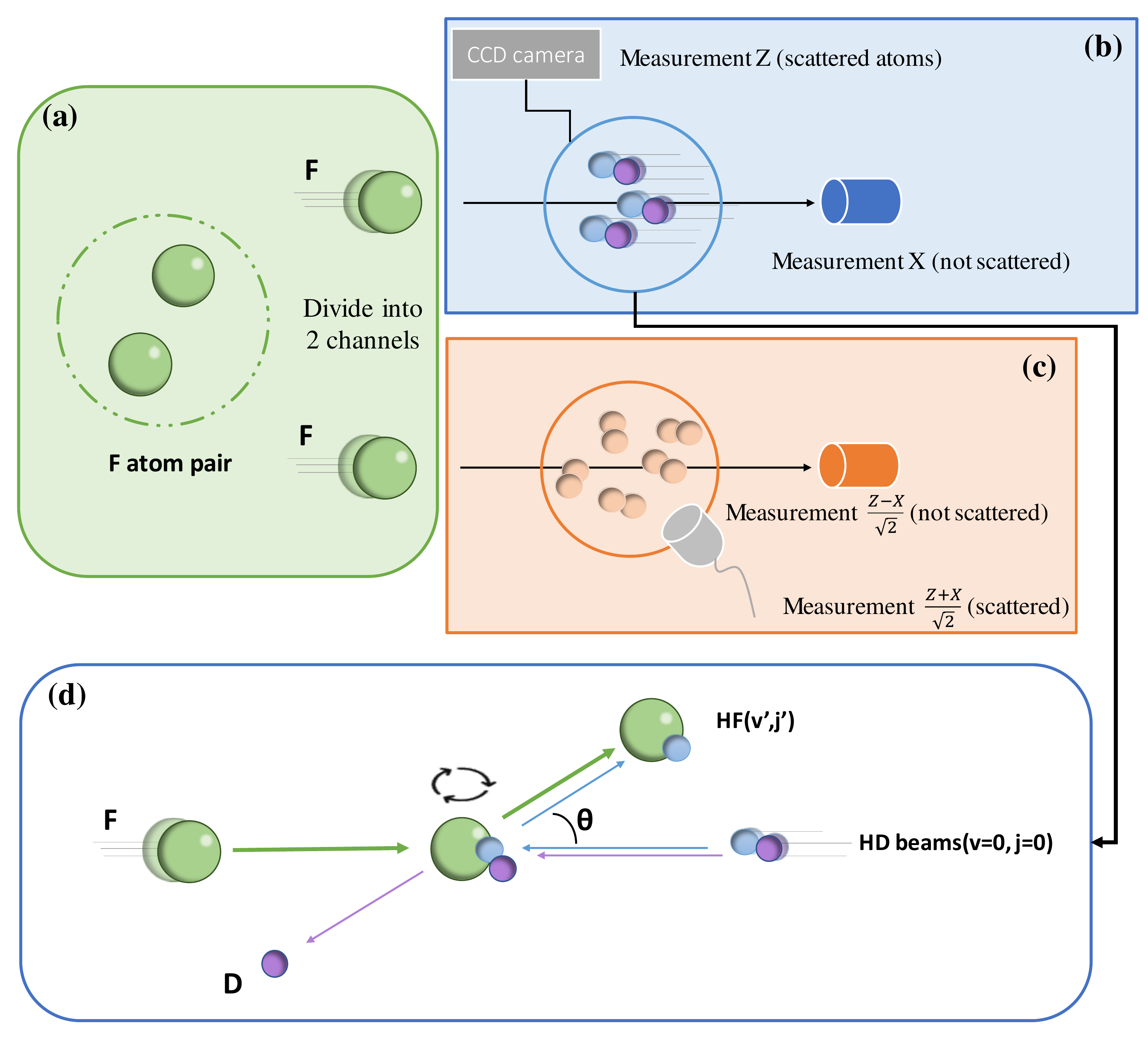}
    \caption{
    {\bf Scheme of the experiment setting.}
    (a)A pair of F atoms are prepared at some certain states initially.
    They are then separated for two channels as shown in (b) and (c).
    (b)In the first channel, the single F atom is sent into the scattering chamber together with a beam of $\ce{HD}(v=0, j=0)$ molecules.
    If the F atom collides with the beam, a charge-coupled device (CCD) camera will capture and record the image, based on which the differential scattering cross-section can be derived.
    The scattering process corresponds to measurement $Z$ with eigenstate $|0\rangle$ and $|1\rangle$.
    The blue cylinder represents is sensor, where the F atom that does not collide with the \ce{HD} beam will be measured under measurement $X$ with eigenstate $|+\rangle$ and $|-\rangle$.
    (c)In the second channel F atom is sent to another scattering chamber.
    The scattering process corresponds to measurement $\frac{Z+X}{\sqrt{2}}$,
    while measurement $\frac{Z-X}{\sqrt{2}}$ will be applied on the atom that avoided collision.
    (d) Sketch of the collision between F atom (green) and $\ce{HD}(v=0, j=0)$ molecule (blue, purple).
    $\theta$ represents the scattering angle between the \ce{HD} beam and the $\ce{HF}(v',j')$ molecule.
    }
    \label{fig_experiment}
\end{figure}

The objective of the experiment is to detect entanglement in the prepared atom pairs, and further study if the  the existence of entanglement would effect the spin-orbit interactions in the transition state.
In the following discussions and simulations, we denote the ground state of \ce{F} (${}^2P_{3/2}$) as $|0\rangle$ state and the excited state (${}^2P_{1/2}$) as $|1\rangle$ state, which are eigenstates of Pauli $Z$ measurement.
Additionally, $|+\rangle=\frac{1}{\sqrt{2}}(|0\rangle+|1\rangle)$ and $|-\rangle=\frac{1}{\sqrt{2}}(|0\rangle-|1\rangle)$ are eigenstates of Pauli $X$ measurement.
Scheme of the experimental design is shown as Fig.(\ref{fig_experiment}).
Initially, two F atoms are prepared at some certain states, such as Bell state for maximum entanglement, mixed states, superposition states or Werner states.
The atom pair will then be separated into two channels as shown in Fig.(\ref{fig_experiment}b,c).
In the first channel the single F atom is sent into the scattering chamber together with a beam of $\ce{HD}(v=0, j=0)$ molecules.
If the F atom collides with the beam, a charge-coupled device (CCD) camera will capture and record the image, based on which the differential scattering cross-section (DCS) can be obtained.
The scattering process corresponds to measurement $Z$ with eigenstate $|0\rangle$ and $|1\rangle$,
while the F atom that does not collide with the \ce{HD} beam will be measured under measurement $X$.
The blue cylinder in Fig.(\ref{fig_experiment}b) represents a sensor applying.
In the other channel, the scattering process corresponds to measurement $\frac{Z+X}{\sqrt{2}}$,
while measurement $\frac{Z-X}{\sqrt{2}}$ will be applied on the atom that avoided the collision.
Fig.(\ref{fig_experiment}d) is a sketch of the reaction $\ce{F}+\ce{HD}\longrightarrow \ce{HF}+\ce{D}$.
The F atom (green) and \ce{HD} molecule (blue and purple) will form a transition state when collision happens and then get scattered into the $\ce{HF}(v',j')$(green and blue) molecule and \ce{D} atom (purple).

\subsection{Necessity for inclusion of Spin-Orbit coupling}
\label{SOC}
The $\ce{F}+\ce{HD}(v=0,j=0)$ reaction shown in Fig.(\ref{fig_experiment}d) will be studied by using a crossed molecular beam (CMB) apparatus with an ion imaging detector\cite{yuan2018direct}.
In the D-atom product velocity map image from $\ce{F}({}^2P_{3/2})+\ce{HD}(v=0,j=0)$ reaction\cite{chen2021quantum}, there are two main ring structures indicating the \ce{HF} products with vibrational states $v'=2$ or $v'=3$.
While the $v'=3$ products are mainly forward scattered, most $v'=2$ products are backward scattered with substantial forward and sideways scatterings.
Particularly, unusual horseshoe shaped structures are detected in the forward-scattering hemisphere around peaked angular distributions of $v'=2$ products in different $j'$ states\cite{chen2021quantum}.
Theoretically, the reactive scattering can be calculated with the coupled two state model or the full six-state model.
The Hamiltonian in Jacobi coordinate could be written as\cite{alexander2000investigation}
\begin{equation}
    H = -\frac{\hbar^2}{2\mu_R}\frac{\partial^2}{\partial R^2}
    -\frac{\hbar^2}{2\mu_r}\frac{\partial^2}{\partial r^2}
    +\frac{L^2}{2\mu_RR^2}+\frac{j^2}{2\mu_rr^2}+V
    \label{Hamitonian}
\end{equation}
where $\mu_R$ the reduced mass between the center of mass of F and \ce{HD} molecule, $R$ is the length of the vector ${\bf R}$ pointing from F atom to the \ce{HD} center of mass, $\mu_r$ is the reduced mass of \ce{HD} and $r$ is the \ce{HD} bond length.
Moreover, $j$ is the is the rotational angular momentum, and $L$ is the operator for
the orbital end-over-end angular momentum of the atom around the center of mass of the diatomic.
In the coupled two state model, the F atoms are treated as structureless particles, where both the electron spin (${\bf s}$) and the electron orbital angular momentum (${\bf l}$) are neglected.
Then  potential energy $V$ under the coupled two state model can be written as\cite{alexander2000investigation}
\begin{equation}
    V = \left(
    \begin{split}
        &V_\Sigma &-\sqrt{2}B\\
        &-\sqrt{2}B &A + V_\Pi\\
    \end{split}
    \right)
    \label{2statePES}
\end{equation}
where the matrix elements $A$ and $B$ describe the usual spin-orbit couplings, $V_\Sigma, V_\Pi$ are determined by the $\emph{ab initio}$ calculations.
In a Cartesian basis, $V_\Sigma=V_{zz}, V_\Pi=(V_{xx}+V_{yy})/2$.

On the other hand, in the full six-state model including the open-shell characteristics of F atoms, $V$ in Eq.(\ref{Hamitonian}) represents the potential energy surface (PES) containing two components $V=V_{el}+V_{so}$.
$V_{el}$ is the adiabatic potential energy surfaces for specific electronic states and nonadiabatic coupling terms, while $V_{so}$ is the electrostatic spin-orbit coupling term, each of which can be written as a $6\times6$ matrix, and can be written as\cite{alexander2000investigation,alexander2002theoretical},
\begin{equation}
    V_{el}=\begin{array}{@{}r@{}c@{}c@{}c@{}c@{}c@{}c@{}l@{}}
   &|\Sigma\rangle &|\Bar{\Sigma}\rangle  &|\Pi_1\rangle &|\Bar{\Pi}_1\rangle &|\Pi_{-1}\rangle &|\Bar{\Pi}_{-1}\rangle \\
   \left.\begin{array}
    {c} |\Sigma\rangle \\ |\Bar{\Sigma}\rangle  \\ |\Pi_1\rangle \\ |\Bar{\Pi}_1\rangle \\ |\Pi_{-1}\rangle \\ |\Bar{\Pi}_{-1}\rangle \end{array}
    \right(
        & \begin{array}{c} V_{\Sigma} \\ 0 \\ -V_1\\ 0 \\ V_1\\ 0\end{array}
        & \begin{array}{c} 0 \\ V_{\Sigma} \\ 0 \\ -V_1 \\ 0\\ V_1\end{array}
        & \begin{array}{c} -V_1 \\ 0 \\ V_{\Pi}\\ 0 \\ V_2\\ 0\end{array}
        & \begin{array}{c} 0 \\ -V_1 \\ 0 \\ V_{\Pi} \\ 0\\ V_2\end{array}
        & \begin{array}{c} V_1 \\ 0 \\ V_2 \\ 0 \\ V_{\Pi}\\ 0\end{array}
        & \begin{array}{c} 0 \\ V_1 \\ 0 \\ V_2 \\ 0\\ V_{\Pi}\end{array}
        & \left)\begin{array}{c} \\ \\ \\ \\ \\ \\\end{array}\right.
  \end{array}
  \label{V_el}
\end{equation}

\begin{equation}
    V_{so}=\begin{array}{@{}r@{}c@{}c@{}c@{}c@{}c@{}c@{}l@{}}
   &|\Sigma\rangle &|\Bar{\Sigma}\rangle  &|\Pi_1\rangle &|\Bar{\Pi}_1\rangle &|\Pi_{-1}\rangle &|\Bar{\Pi}_{-1}\rangle \\
   \left.\begin{array}
    {c} |\Sigma\rangle \\ |\Bar{\Sigma}\rangle  \\ |\Pi_1\rangle \\ |\Bar{\Pi}_1\rangle \\ |\Pi_{-1}\rangle \\ |\Bar{\Pi}_{-1}\rangle \end{array}
    \right(
        & \begin{array}{c} 0 \\ 0 \\ 0\\ -\sqrt{2}B \\ 0\\ 0\end{array}
        & \begin{array}{c} 0 \\ 0 \\ 0 \\ 0 \\-\sqrt{2}B \\ 0\end{array}
        & \begin{array}{c} 0 \\ 0 \\ -A \\ 0 \\ 0\\ 0\end{array}
        & \begin{array}{c} -\sqrt{2}B \\ 0 \\ 0 \\ A \\ 0\\ 0\end{array}
        & \begin{array}{c} 0 \\  -\sqrt{2}B \\ 0 \\ 0 \\ A \\ 0\end{array}
        & \begin{array}{c} 0 \\ 0 \\ 0 \\ 0 \\ 0\\ -A\end{array}
        & \left)\begin{array}{c} \\ \\ \\ \\ \\ \\\end{array}\right.
  \end{array}
  \label{V_so}
\end{equation}
Similarly to $V_\Sigma, V_\Pi$, $V_1, V_2$ are also determined by the $\emph{ab initio}$ calculation as $V_1=V_{xz}/\sqrt{2}$, $V_2=(V_{yy}-V_{xx})/2$.

\begin{figure}[ht]
    \centering
    \begin{subfigure}[t]{0.4\textwidth}
        \centering
        \includegraphics[width=\textwidth]{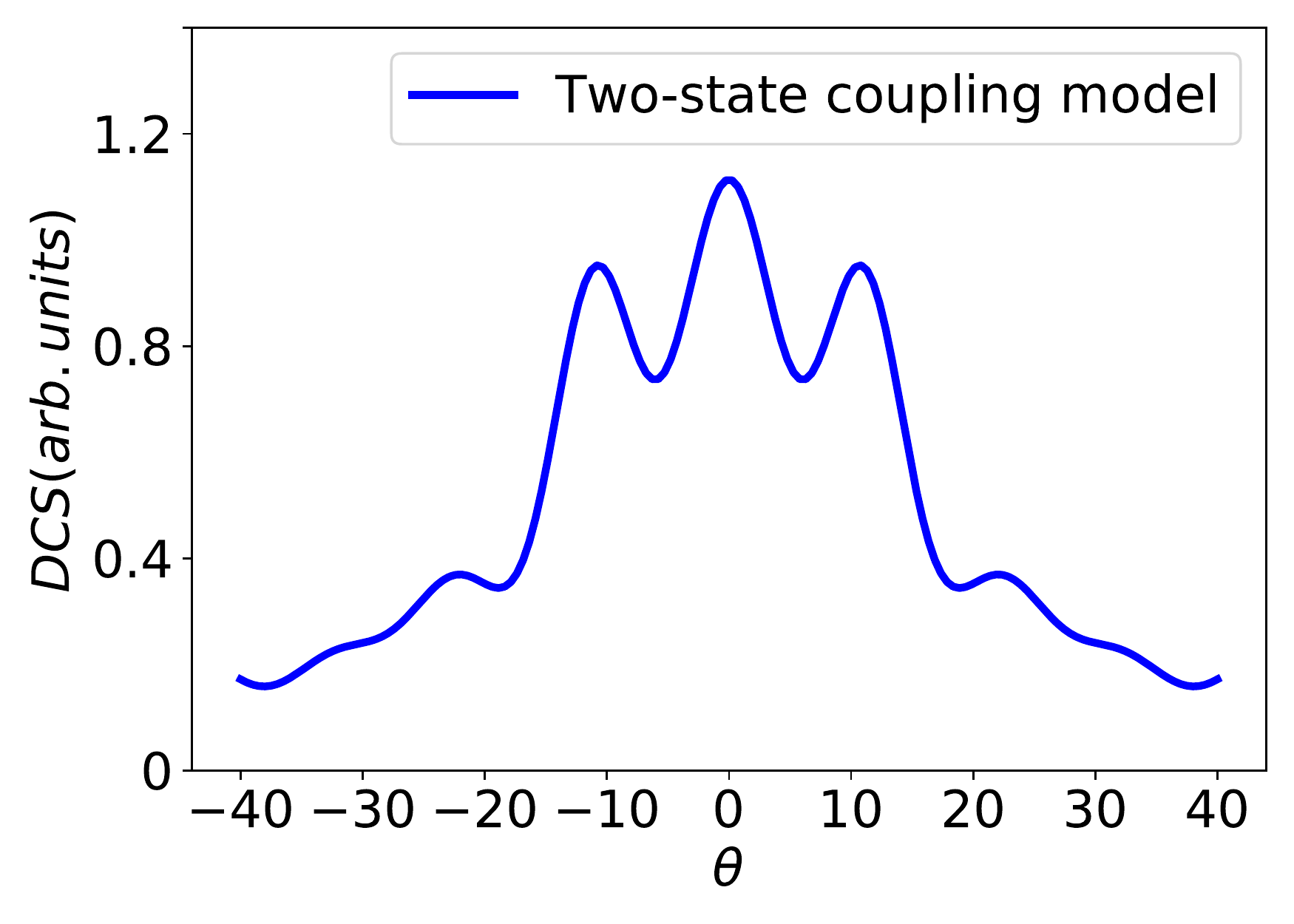}
        \caption{}
        \label{horseshoe2}
    \end{subfigure}
    \begin{subfigure}[t]{0.4\textwidth}
        \centering
        \includegraphics[width=\textwidth]{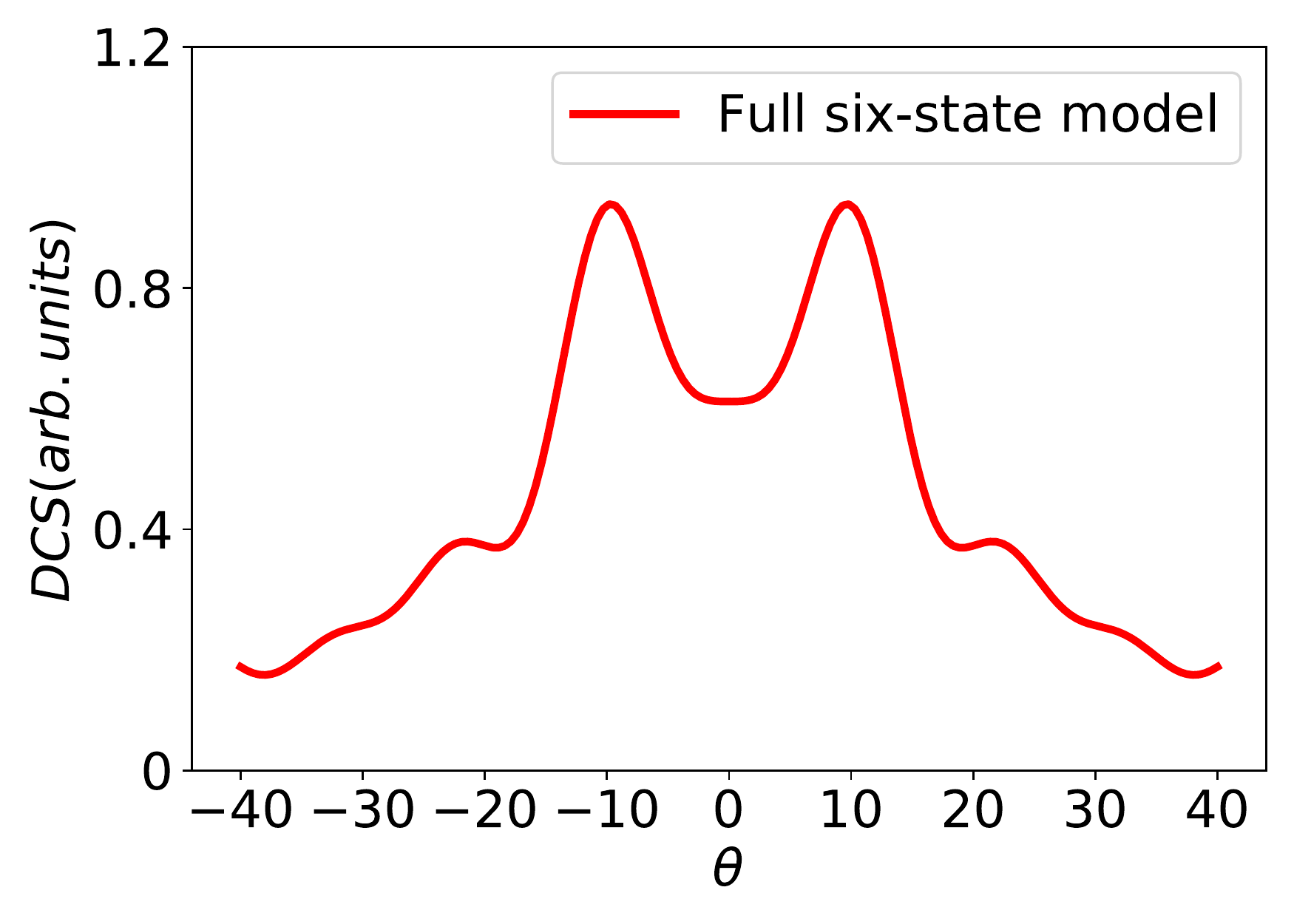}
        \caption{}
        \label{horseshoe6}
    \end{subfigure}
    \caption{
    {\bf Theoretically calculated DCSs for product $\ce{HF}(v'=2, j'=5)$ around the forward-scattering direction in the $\ce{F}+\ce{HD}$ reaction}
    (a)Theoretically calculated DCS based on two-state coupling model, where the F atoms are treated as structureless particles.
    Three peaks arouse around the forward-scattering direction.
    (b)Theoretically calculated DCS based on full six-state model, where full spin-orbit effects are included.
    The theoretical DCSs fit well with the horse shoe shape pattern observed in experiments\cite{chen2021quantum}.
    In both figures, $\theta$ is the scattering angle, and collision energy is around $2.10 kcal/mol$.
    }
    \label{fig_circuit}
\end{figure}

Theoretically calculated DCS for product $\ce{HF}(v'=2, j'=5)$\cite{chen2021quantum} with these two models are shown in figure(\ref{horseshoe2},\ref{horseshoe6}).
Two-state coupling model leads to three peaks around the forward-scattering direction, instead of the horseshoe shape pattern.
The DCS from full six-state model fits the experiment results very well, while the two-state coupling model can not explain the unusual horseshoe shape signature, indicating that the full spin-orbit effects is essential to explain the $\ce{F}+\ce{HD}$ reaction\cite{chen2021quantum}.
Thus, the horseshoe pattern in $\ce{F}+\ce{HD}$ reaction is reliable evidence for the spin-orbit characteristic.

\subsection{Detection of Entanglement}
\label{Entanglement_det}
As shown in Fig.(\ref{fig_experiment}b,c), there are two scattering processes  and two direct measurements on the unscattered particles in each channel. Two scattering processes correspond to the measurement $Z$ and measurement $\frac{Z+X}{\sqrt{2}}$, while measurement $X$ and $\frac{Z-X}{\sqrt{2}}$ are applied on the unscattered particles.
All four measurements together implement a Clauser-Horne-Shimony-Holt (CHSH) experiment\cite{aspect1982experimental} detecting the existence of entanglement.
Generally, the $CHSH$ inequality can be written as
\begin{equation}
    \left|
    E(r_1, t_1)-E(r_1, t_2)+E(r_2, t_1)+E(r_2, t_2)
    \right|
    \leq2
    \label{ineq_chsh}
\end{equation}
where the variables $r_i$ and $t_i$ $\forall\:\: i$ encodes the respective outcomes of the two measurement observables $(\tilde{r},\tilde{t})$. $E(r_i,t_i)$ is a measure of quantum correlation in the results obtained from a pair of such outcomes. In the standard Bell test or CHSH experiment, all the four measurements lead to binary values, i.e $\{r_i, t_i\} \in \{+,-\}$. The measure of quantum correlation $E(r_i,t_i)$ in such a case is definable as
\begin{equation}
    E(r,t)=
    \frac{N_{++}-N_{+-}-N_{-+}+N_{--}}
    {N_{++}+N_{+-}+N_{-+}+N_{--}}
    \label{ert}
\end{equation}
where $N_{++}$ represents the number of particle pairs yielding result $+$ in both the measurement of observable $\tilde{r}$ of channel I and measurement of observable $\tilde{t}$ of channel II. The four $E(r_i,t_i)$ terms form the test statistics parameter $S =E(r_1, t_1)-E(r_1, t_2)+E(r_2, t_1)+E(r_2, t_2)$.

However, in the experiment shown in Fig.(\ref{fig_experiment}), chemical reactions are included, which leads to continuous measurement outcomes.
Thus, Eq.(\ref{ert}), the standard definition of quantum correlations in Bell test, no longer works in such a scattering process.
For chemical reactions, instead, the  generalized CHSH inequality for continuous variables\cite{li2019entanglement} is required. We denote $M({\bf x_1},{\bf x_2} |\rho,\tilde{r},\tilde{t}) d{\bf x_2}d{\bf x_1}$ as the probability to get a measurement outcome  within volume d${\bf x_1}$ centered at ${\bf x_1}$ under observable $\tilde{r}$ and within volume d${\bf x_2}$ centered at ${\bf x_2}$ under observable $\tilde{t}$ for a particle pair with density matrix $\rho$.
We always have $M({\bf x_1},{\bf x_2}|\rho,\tilde{r},\tilde{t})\geq0$ and $\int d{\bf x_1}d{\bf x_2}M({\bf x_1}, {\bf x_2}|\rho,\tilde{r},\tilde{t})=1$.
Similarly, we denote $M({\bf x_1}|\rho^\prime, \tilde{r})$ as the marginal probability density to get measurement result in the neighborhood of ${\bf x_1}$ under measurement observable $\tilde{r}$
irrespective of the outcome of the observable $\tilde{t}$ for a single particle with density matrix $\rho^\prime$.  Let us define an auxiliary function for the outcome  measurement $\tilde{r}$ and $\tilde{t}$ as\cite{li2019entanglement}
\begin{equation}
    V ({\bf x_1}, {\bf x_2}|\tilde{r}, \tilde{t}) = 
    [v({\bf x_1}|+,\tilde{r}) - v({\bf x_1}|-,\tilde{r})]\cdot
    [v({\bf x_2}|+,\tilde{t}) - v({\bf x_2}|-,\tilde{t})]
\end{equation}
and $v({\bf x}|\pm,\tilde{r})$ are defined as
\begin{equation}
    v({\bf x}|\pm,\tilde{r})=
    \frac{M({\bf x}||\phi_r^\pm\rangle\ \langle\phi_r^\pm|, \tilde{r})
    -\int d{\bf x}M({\bf x}||\phi_r^\pm\rangle\ \langle\phi_r^\pm|, \tilde{r})
    M({\bf x}||\phi_r^\mp\rangle\ \langle\phi_r^\mp|, \tilde{r} )}
    {
    \int d{\bf x}M({\bf x}||\phi_r^\pm\rangle\ \langle\phi_r^\pm|, \tilde{r} )
    M({\bf x}||\phi_r^\pm\rangle\ \langle\phi_r^\pm|, \tilde{r})
    -
    \int d{\bf x}M({\bf x}||\phi_r^\pm\rangle\ \langle\phi_r^\pm|, \tilde{r} )
    M({\bf x}||\phi_r^\mp\rangle\ \langle\phi_r^\mp|, \tilde{r})
    }
\end{equation}
where $|\phi_r^\pm\rangle$ represent the two eigenstates corresponding to values $\{+,-\}$ in measurement of $\tilde{r}$. Similar considerations applies to $\tilde{t}$-measurement as well.
Then we get the generalized CHSH inequality for continuous variables taking the following form
\begin{equation}
    \begin{split}
    \left|
    \int d{\bf x_1}d{\bf x_2} M({\bf x_1},{\bf x_2}|\rho, \tilde{r_1}, \tilde{t_1})
    V ({\bf x_1}, {\bf x_2}|\tilde{r_1}, \tilde{t_1})
    -\int d{\bf x_1}d{\bf x_2} M({\bf x_1}, {\bf x_2}|\rho, \tilde{r_1}, \tilde{t_2})
    V ({\bf x_1}, {\bf x_2}|\tilde{r_1}, \tilde{t_2})
    \right.
    \\
    \left.
    +\int d{\bf x_1}d{\bf x_2} M({\bf x_1},{\bf x_2}|\rho, \tilde{r_2}, \tilde{t_1})
    V ({\bf x_1}, {\bf x_2}|\tilde{r_2}, \tilde{t_1})
    +\int d{\bf x_1}d{\bf x_2} M({\bf x_1},{\bf x_2}|\rho, \tilde{r_2}, \tilde{t_2})
    V ({\bf x_1}, {\bf x_2}|\tilde{r_2}, \tilde{t_2})
    \right|
    \leq2
    \end{split}
    \label{chsh_continuous}
\end{equation}
More discussion about the detection of entanglement in chemical reactions can be found in our recent work\cite{li2019entanglement}.

\section{Numerical Simulation}
\label{Num_sim}

In this section, we predict the possible results for the experiment shown in Fig.(\ref{fig_experiment}).
Consider that the F atom pairs in the experiment shown in Fig.(\ref{fig_experiment}) are prepared initially at bipartite Werner state\cite{werner1989quantum}, whose density matrix can be written as
\begin{equation}
    \rho_W=
    \begin{pmatrix}
    \frac{1-p}{4} &0 &0 &0\\
    0  &\frac{1+p}{4} &\frac{p}{2} &0\\
    0  &\frac{p}{2} &\frac{1+p}{4} &0\\
    0 &0 &0 &\frac{1-p}{4}
    \end{pmatrix}
\end{equation}
where $0\leq p\leq 1$.
When $p=1$, the Werner state yields a pure Bell state, with maximum entanglement in bipartite systems,
while if $p=0$, the Werner state degenerates into a uniformly mixed state whose density matrix is just $\frac{I}{4}$, with $I$ being the identity matrix.

\begin{figure}[ht]
    \centering
    \begin{subfigure}[t]{0.4\textwidth}
        \centering
        \includegraphics[width=\textwidth]{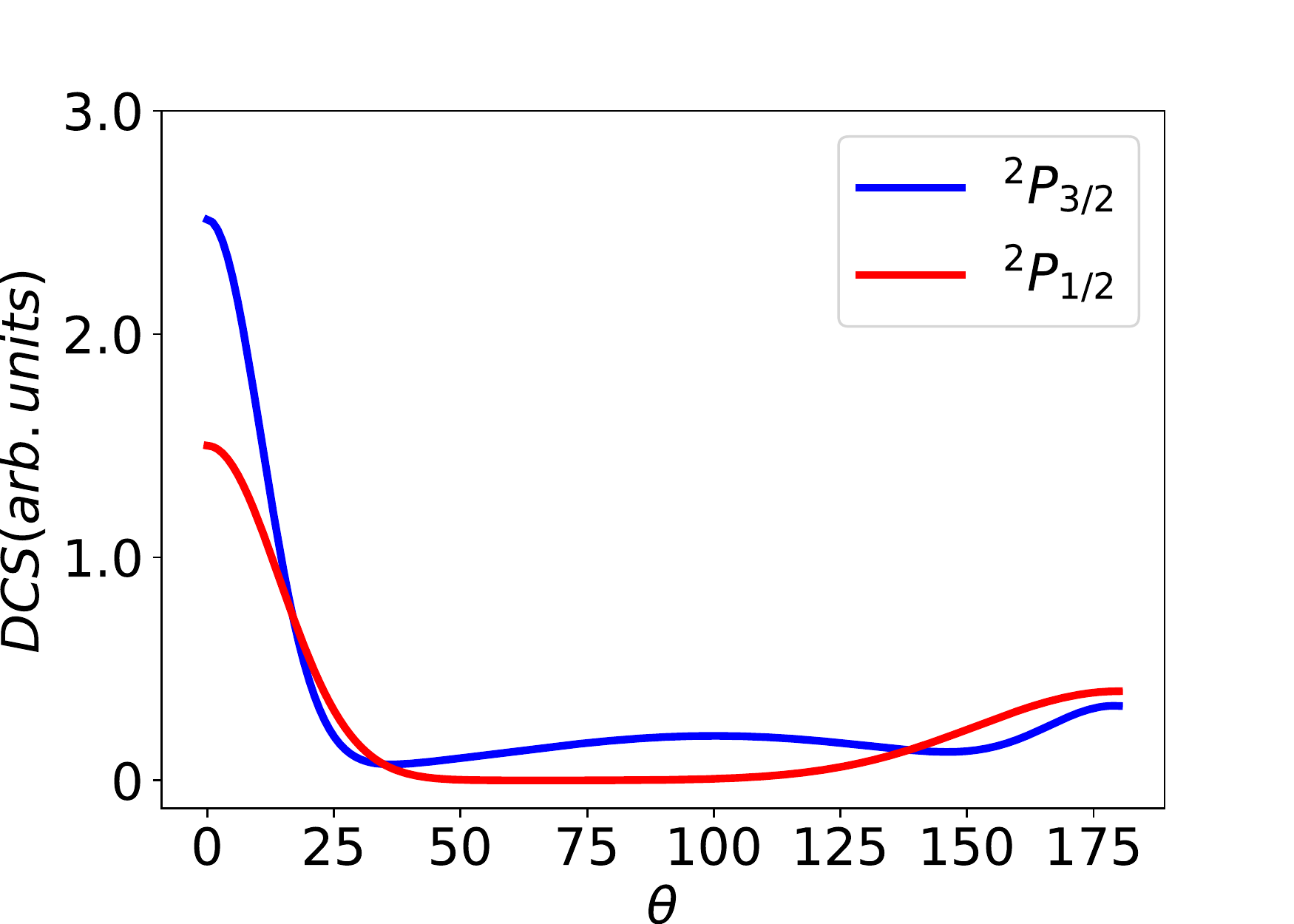}
        \caption{}
        \label{channel1}
    \end{subfigure}
    \begin{subfigure}[t]{0.4\textwidth}
        \centering
        \includegraphics[width=\textwidth]{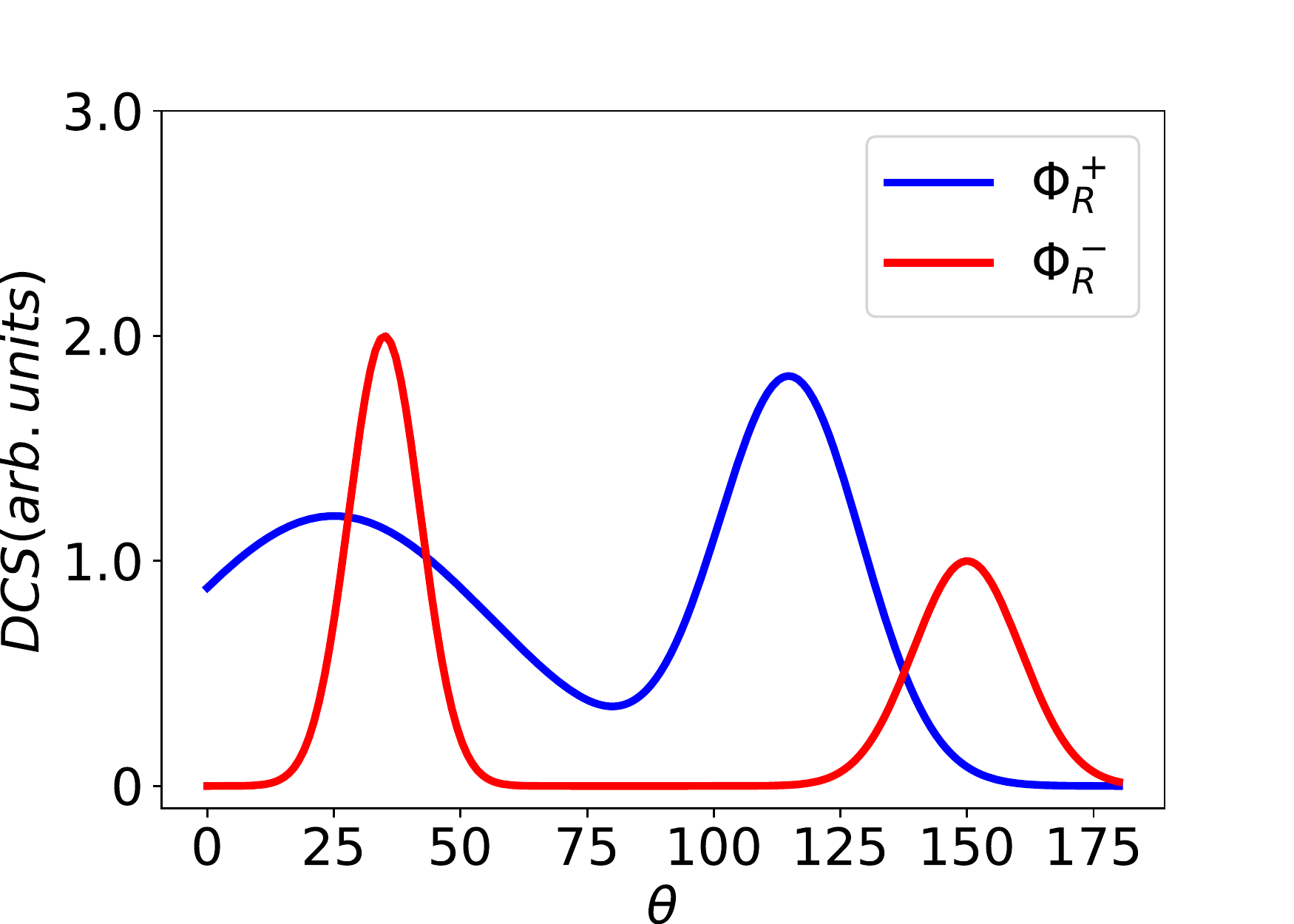}
        \caption{}
        \label{channel2}
    \end{subfigure}
    \\
    \centering
    \begin{subfigure}[t]{0.8\textwidth}
        \centering
        \includegraphics[width=\textwidth]{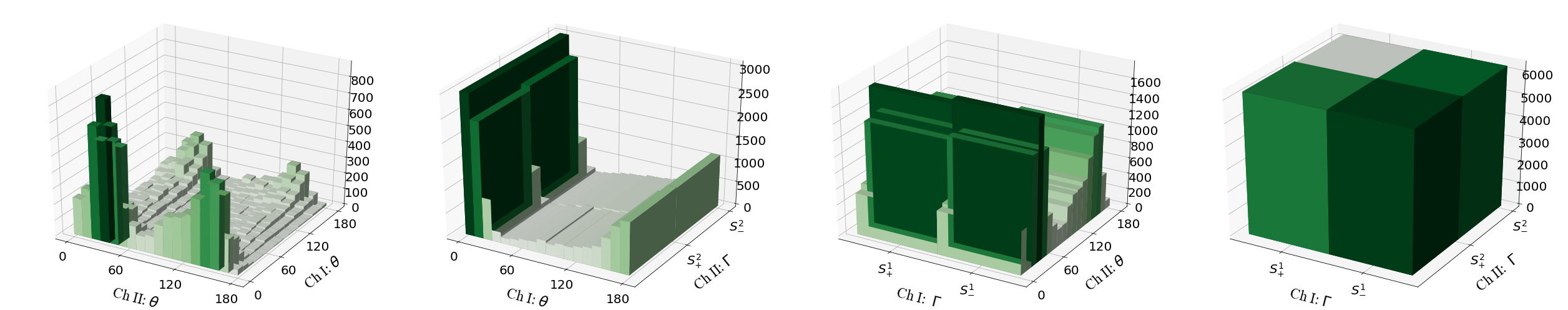}
        \caption{}
        \label{fig_p0}
    \end{subfigure}
    \\
    \centering
    \begin{subfigure}[t]{0.8\textwidth}
        \centering
        \includegraphics[width=\textwidth]{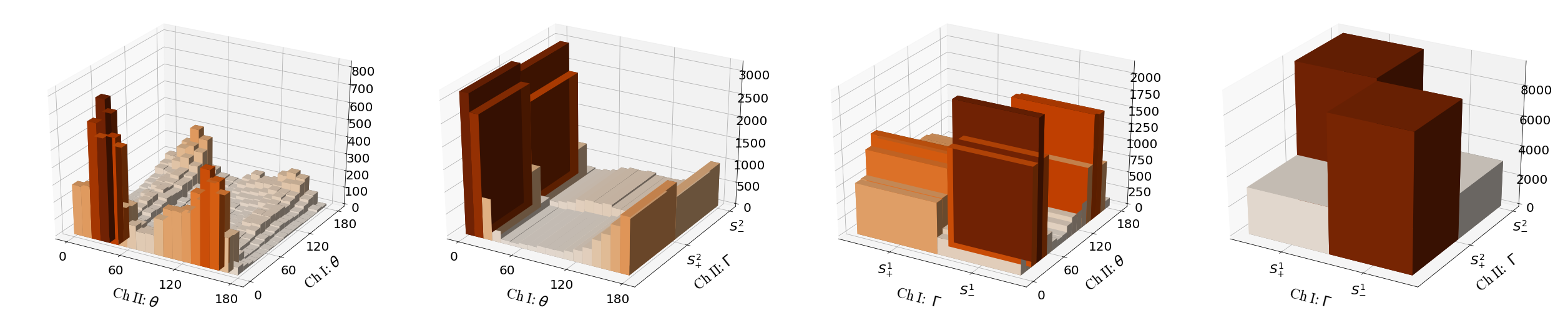}
        \caption{}
        \label{fig_p0707}
    \end{subfigure}
    \\
    \centering
    \begin{subfigure}[t]{0.8\textwidth}
        \centering
        \includegraphics[width=\textwidth]{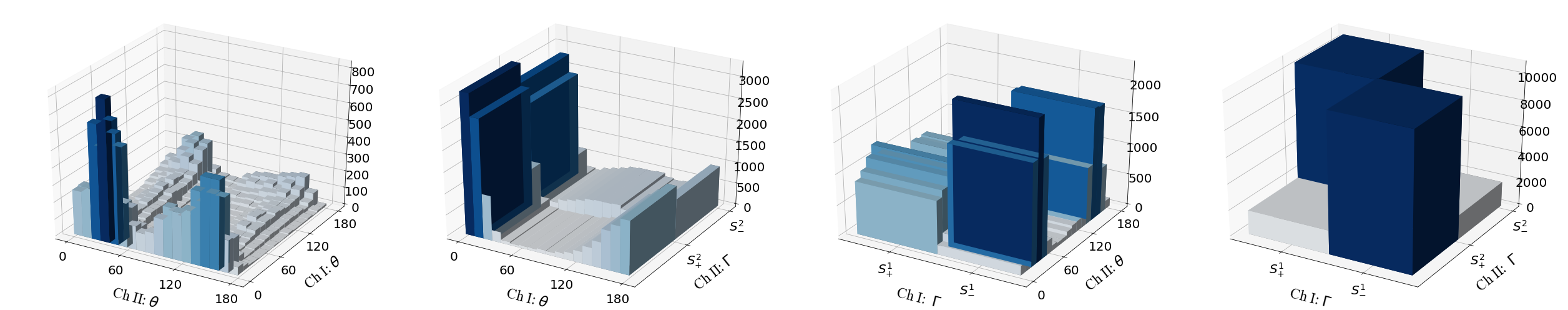}
        \caption{}
        \label{fig_p1}
    \end{subfigure}
    \caption{
    {\bf DCSs for the reactions and numerical simulations for various Werner states.}
    (a)Scheme of the DCS for the $\ce{F}({}^2P_{1/2,3/2})+\ce{HD}(v=0,j=0)$ reaction, only the $\ce{HF}(v'=3)$ product is included.
    (b)DCS of $|\Phi_R^{\pm}\rangle$ in the reaction corresponding to measurement $\frac{Z+X}{\sqrt{2}}$,
    where $|\Phi_R^{\pm}\rangle$ are eigenstates of $\frac{Z+X}{\sqrt{2}}$.
    (c,d,e)Numerical simulations Werner states with various $p$ values $p=0,0.707, 1$.
    Histograms represent the count of F atom pairs with certain measurement results, 
    where $\theta$ is the scattering angle of product $\ce{HF}(v'=3)$, and $\Gamma$ is the direct measurement result with two possible values denoted as $S_{\pm}$.
    In the first column we show the simulation results that both the F atoms are scattered in two channels.
    In the second one we show results that F atom is scattered in channel I but not scattered in channel II.
    The third column contains simulation results that F atom is not scattered in channel I but scattered in channel II.
    In the last column we show the simulation results that neither the F atoms are scattered in two channels.
    }
    \label{fig_simulation}
\end{figure}

We simulated the experimental results as shown in Fig.(\ref{fig_simulation}).
In the first channel shown in Fig.(\ref{fig_experiment}b),  the F atom goes into scattering chamber together with the \ce{HD} beams.
The collision energy is set around $2.10 kcal/mol$, where the horseshoe shape pattern (see Fig.\ref{horseshoe6}) is observed in certain product distribution.
Fig.(\ref{channel1}) is a scheme of the DCS for the $\ce{F}({}^2P_{1/2,3/2})+\ce{HD}(v=0,j=0)$ reaction. The only product included in this data is the $\ce{HF}(v'=3)$. The data is digitized from \cite{lee2004resonance} by fitting the relevant peaks to gaussian functions. The two DCS curves in this plot represents the two outcomes of the Z-measurement. On the other hand, in another channel shown in Fig.(\ref{fig_experiment}c) the scattering process corresponds to measurement by $\frac{Z+X}{\sqrt{2}}$ operator.
For simplicity, here we assume that the DCS for $|\Phi_R^{\pm}\rangle$ are well described as shown in Fig.(\ref{channel2}),
where $|\Phi_R^{\pm}\rangle$ are eigenstates of $\frac{Z+X}{\sqrt{2}}$. Due to lack of experimental results, the corresponding DCS in Fig.(\ref{channel2}) are randomly generated patterns. It must be emphasized that the overall assertion in this report will remain unchanged for any other pattern as well as long as the DCS curves for the $|\Phi_R^{\pm}\rangle$ have sufficiently different support as is the case in Fig.(\ref{channel2}).

Numerical simulations predict the experimental results as shown in Fig.(\ref{fig_p0},\ref{fig_p0707},\ref{fig_p1}), for Werner states with various $p$ values $p=0,0.707, 1$. The variable 
$\theta$ in these subfigures is the scattering angle of product $\ce{HF}(v'=3)$, and $\Gamma$ is the direct measurement result with two possible values denoted as $S_{\pm}$ which are either the $Z$-measurement result in chamber I with observed value of $\pm$1 or $\frac{Z+X}{\sqrt{2}}$ value in chamber II with observed value of $\pm$1. The vertical axis in each subfigure represent the frequency of F atom pairs with a certain measurement results . For example in the 1st column of Fig.(\ref{fig_p0}), the vertical axis shows the count of $\ce{F}$ atom pairs, one of which have been detected by direct $Z$-measurement in chamber I and have also scattered 
a $\ce{HF}(v'=3)$ within a neighborhood of $\theta_1$ (say) while the other have been detected in chamber II by $\frac{Z+X}{\sqrt{2}}$ and have concomitantly scattered 
a $\ce{HF}(v'=3)$ within a neighborhood of $\theta_2$ (say). The second column shows the count of $\ce{F}$ atom pairs one of which have scattered a $\ce{HF}(v'=3)$ in channel I within a given neighborhood of $\theta$ and the other have resulted in a $\frac{Z+X}{\sqrt{2}}$ measurement of $\pm$ 1. Similar interpretation can be constructed for the horizontal and vertical axes of the third and fourth column in the said figure. All axes in columns of Fig.(\ref{fig_p0707}) and Fig.(\ref{fig_p1}) have similar meaning as these but simulated for Werner states of the initial $\ce{F}$ atom pairs with different $p$ values. For the case of each of the $p$ values, we simulated the experiment 100,000 times, and assume that the probability of single F atoms to be scattered is $0.5$, regardless to its initial states. For $p=0.707$ which is the onset of an entanglement and $p=1.0$ which corresponds to a Bell state we do see the expected anti-correlation in Fig.(\ref{fig_p0707}) and Fig.(\ref{fig_p1}) especially in 4th column. Such features are absent for $p=0$ which corresponds to a maximally mixed state. Even though Werner states with various $p$ values might lead to such extreme differences,
it could always be challenging to read out these features directly from these counts. Alternatively, we can estimate the test statistics formed by quantum correlations with the CHSH inequality for continuous variables shown in Eq.(\ref{chsh_continuous}) which we discuss next. 

\begin{figure}[ht]
    \centering
    \includegraphics[width=0.6\textwidth]{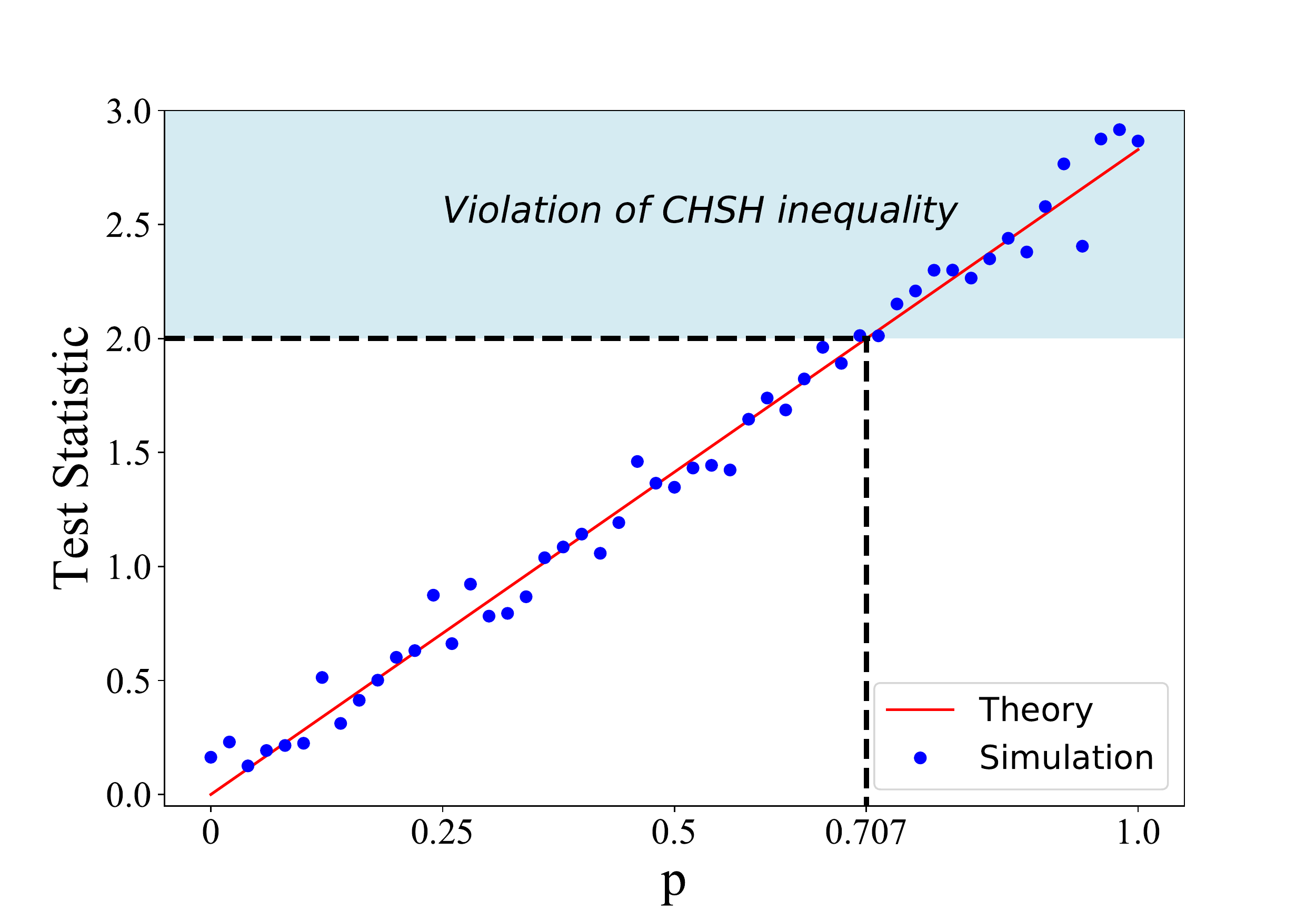}
    \caption{
    {\bf Test statistics for various $p$ values.}
    The red line indicates the theoretical prediction, and the blue dots represent simulation results.
    When $p\geq1/\sqrt{2}$, it is expected to observe violation of the CHSH inequality,
    which guarantees the existence of entanglement.
    The light blue area indicates the violation of the CHSH inequality.
    }
    \label{fig_inequality}
\end{figure}

Alternatively, we can estimate the test statistics formed by quantum correlations with the CHSH inequality for continuous variables shown in Eq.(\ref{chsh_continuous}).
Test statistics for various $p$ values are shown in Fig.(\ref{fig_inequality}), where the red line indicates the theoretical prediction, and the blue dots represent simulation results.
When $p\geq 0.707$, it is expected to observe violation of the CHSH inequality,
in other words, the test statistic is greater than $2$,
which guarantees the existence of entanglement.

In brief,  the existence of entanglement can be detected with the generalized CHSH inequality Eq.(\ref{chsh_continuous}), while the special patterns in DSC indicates the necessity for the inclusion of spin-orbit characteristics (see Fig.\ref{horseshoe6}).
With the experimental implementation shown in Fig.(\ref{fig_experiment}), the test statistic can be estimated based on the collected data from two channels with four measurements and the frequency of $\ce{F}$ atom pairs detected satisfying the possible outcomes of the measurements can be plotted as shown in Fig.\ref{fig_p0} (or 
Fig.\ref{fig_p0707} , Fig.\ref{fig_p1}). Violation of the generalized CHSH inequality in the count statistics guarantees the existence of entanglement.
On the other hand, we need to focus on the DCSs of certain products. Observation of the horseshoe shape patterns like Fig.(\ref{horseshoe6}) in product $\ce{HF}(v'=2, j'=5)$ indicate the full spin-orbit characteristics.
Observation of the horseshoe shape patterns, along with violation of the CHSH inequality, indicate that the existence of entanglement between the initial F atom pairs has little influence on the spin-orbit characteristics, so that the six-state still works as before.
In contrast, another possible result is that the CHSH inequality is violated, meanwhile there is no horseshoe shape patterns in the forward-scattering of $v'=2$ products.
If so, we need to admit that the potential energy surface (PES) changes with the entanglement between F atom pairs. For example, patterns like Fig.(\ref{horseshoe2}) in the forward-scattering suggest that the PES reduce to Eq.(\ref{2statePES}), which can be described with the simple two state model.
Additionally, entanglement might be vulnerable under the collisions. If so we need to apply lower collision energy, ensuring that entanglement will not be broken in the reactions.

\section{Conclusion}
In this paper we propose an experiment to study the possible statistical correlation between entanglement and features associated with spin-orbit coupling,
where both the existence of entanglement and the spin-orbit characteristics can be detected simultaneously.
Particularly we propose an implementation of the experiment based on the $\ce{F}+\ce{HD}$ reaction.
The existence of entanglement can be guaranteed from the violation of generalized CHSH inequality for continuous variables which can be ascertained from the frequency of $\ce{F}$ atom pair distribution, while the special patterns in DSC from the product $\ce{HF}$ distribution would mandate the necessity for the inclusion of spin-orbit characteristics.
We further numerically simulated the possible experimental results , pointing out the key features for various realizable outputs. under the assumption that the DCS curves for two of the possible direct measurements $\frac{Z \pm X}{\sqrt{2}}$ have nearly disjointed supports (see Fig.\ref{channel2}). With advancement in experimental controls, the marriage of such studies which hope to illustrate hidden correlations between initial reactant molecules (atoms) and known geometrical features of product distribution can be undertaken and even possibly exploited as a formidable resource. 

\section*{Acknowledgement}
\label{Acknowledgement}
We acknowledge funding by the U.S. Department of Energy (Office of Basic Energy Sciences) under Award No. DE-SC0019215, and the  National Science Foundation under Award No. 1955907.

\bibliography{ref}

\begin{thebibliography}{10}

\bibitem{whaley2014quantum}
Sabre Kais, K~Birgitta Whaley, Aaron~R Dinner, and Stuart~A Rice.
\newblock {\em Quantum information and computation for chemistry}.
\newblock John Wiley \& Sons, 2014.

\bibitem{preskill2021quantum}
John Preskill.
\newblock Quantum computing 40 years later.
\newblock {\em arXiv preprint arXiv:2106.10522}, 2021.

\bibitem{ursin2007entanglement}
Rupert Ursin, F~Tiefenbacher, T~Schmitt-Manderbach, H~Weier, Thomas Scheidl,
  M~Lindenthal, B~Blauensteiner, T~Jennewein, J~Perdigues, P~Trojek, et~al.
\newblock Entanglement-based quantum communication over 144 km.
\newblock {\em Nature physics}, 3(7):481--486, 2007.

\bibitem{zhang2017quantum}
Wei Zhang, Dong-Sheng Ding, Yu-Bo Sheng, Lan Zhou, Bao-Sen Shi, and Guang-Can
  Guo.
\newblock Quantum secure direct communication with quantum memory.
\newblock {\em Physical review letters}, 118(22):220501, 2017.

\bibitem{degen2017quantum}
Christian~L Degen, F~Reinhard, and Paola Cappellaro.
\newblock Quantum sensing.
\newblock {\em Reviews of modern physics}, 89(3):035002, 2017.

\bibitem{pirandola2018advances}
Stefano Pirandola, B~Roy Bardhan, Tobias Gehring, Christian Weedbrook, and Seth
  Lloyd.
\newblock Advances in photonic quantum sensing.
\newblock {\em Nature Photonics}, 12(12):724--733, 2018.

\bibitem{giovannetti2006quantum}
Vittorio Giovannetti, Seth Lloyd, and Lorenzo Maccone.
\newblock Quantum metrology.
\newblock {\em Physical review letters}, 96(1):010401, 2006.

\bibitem{karra2016prospects}
Mallikarjun Karra, Ketan Sharma, Bretislav Friedrich, Sabre Kais, and Dudley
  Herschbach.
\newblock Prospects for quantum computing with an array of ultracold polar
  paramagnetic molecules.
\newblock {\em The Journal of chemical physics}, 144(9):094301, 2016.

\bibitem{cao2019quantum}
Yudong Cao, Jonathan Romero, Jonathan~P Olson, Matthias Degroote, Peter~D
  Johnson, M{\'a}ria Kieferov{\'a}, Ian~D Kivlichan, Tim Menke, Borja
  Peropadre, Nicolas~PD Sawaya, et~al.
\newblock Quantum chemistry in the age of quantum computing.
\newblock {\em Chemical reviews}, 119(19):10856--10915, 2019.

\bibitem{king2021scaling}
Andrew~D King, Jack Raymond, Trevor Lanting, Sergei~V Isakov, Masoud Mohseni,
  Gabriel Poulin-Lamarre, Sara Ejtemaee, William Bernoudy, Isil Ozfidan,
  Anatoly~Yu Smirnov, et~al.
\newblock Scaling advantage over path-integral monte carlo in quantum
  simulation of geometrically frustrated magnets.
\newblock {\em Nature communications}, 12(1):1--6, 2021.

\bibitem{kais2007entanglement}
Sabre Kais.
\newblock Entanglement, electron correlation, and density matrices.
\newblock {\em Advances in Chemical Physics}, 134:493, 2007.

\bibitem{bayer2001coupling}
M~Bayer, Pawel Hawrylak, K~Hinzer, S~Fafard, Marek Korkusinski, ZR~Wasilewski,
  O~Stern, and A~Forchel.
\newblock Coupling and entangling of quantum states in quantum dot molecules.
\newblock {\em Science}, 291(5503):451--453, 2001.

\bibitem{shankar2013autonomously}
Shyam Shankar, Michael Hatridge, Zaki Leghtas, KM~Sliwa, Aniruth Narla, Uri
  Vool, Steven~M Girvin, Luigi Frunzio, Mazyar Mirrahimi, and Michel~H Devoret.
\newblock Autonomously stabilized entanglement between two superconducting
  quantum bits.
\newblock {\em Nature}, 504(7480):419--422, 2013.

\bibitem{wang2016experimental}
Xi-Lin Wang, Luo-Kan Chen, Wei Li, H-L Huang, Chang Liu, Chao Chen, Y-H Luo,
  Z-E Su, Dian Wu, Z-D Li, et~al.
\newblock Experimental ten-photon entanglement.
\newblock {\em Physical review letters}, 117(21):210502, 2016.

\bibitem{huang2005entanglement}
Zhen Huang and Sabre Kais.
\newblock Entanglement as measure of electron--electron correlation in quantum
  chemistry calculations.
\newblock {\em Chemical physics letters}, 413(1-3):1--5, 2005.

\bibitem{lin2020quantum}
Yiheng Lin, David~R Leibrandt, Dietrich Leibfried, and Chin-wen Chou.
\newblock Quantum entanglement between an atom and a molecule.
\newblock {\em Nature}, 581(7808):273--277, 2020.

\bibitem{huang2006entanglement}
Zhen Huang and Sabre Kais.
\newblock Entanglement evolution of one-dimensional spin systems in external
  magnetic fields.
\newblock {\em Physical Review A}, 73(2):022339, 2006.

\bibitem{oh2008entanglement}
Sangchul Oh, Zhen Huang, Uri Peskin, and Sabre Kais.
\newblock Entanglement, berry phases, and level crossings for the atomic
  breit-rabi hamiltonian.
\newblock {\em Physical Review A}, 78(6):062106, 2008.

\bibitem{zhu2012multipartite}
Jing Zhu, Sabre Kais, Al{\'a}n Aspuru-Guzik, Sam Rodriques, Ben Brock, and
  Peter~J Love.
\newblock Multipartite quantum entanglement evolution in photosynthetic
  complexes.
\newblock {\em The Journal of chemical physics}, 137(7):074112, 2012.

\bibitem{pauls2013quantum}
James~A Pauls, Yiteng Zhang, Gennady~P Berman, and Sabre Kais.
\newblock Quantum coherence and entanglement in the avian compass.
\newblock {\em Physical review E}, 87(6):062704, 2013.

\bibitem{blasing2018observation}
David~B Blasing, Jes{\'u}s P{\'e}rez-R{\'\i}os, Yangqian Yan, Sourav Dutta,
  Chuan-Hsun Li, Qi~Zhou, and Yong~P Chen.
\newblock Observation of quantum interference and coherent control in a
  photochemical reaction.
\newblock {\em Physical review letters}, 121(7):073202, 2018.

\bibitem{jambrina2015quantum}
Pablo~G Jambrina, Diego Herr{\'a}ez-Aguilar, F~Javier Aoiz, Mahima Sneha,
  Justinas Jankunas, and Richard~N Zare.
\newblock Quantum interference between h+ d 2 quasiclassical reaction
  mechanisms.
\newblock {\em Nature chemistry}, 7(8):661--667, 2015.

\bibitem{kale2021constructive}
Sumit~Suresh Kale, Yong~P. Chen, and Sabre Kais.
\newblock Constructive quantum interference in a photo-chemical reaction of
  $^{87}$rb bose einstein condensate, 2021.

\bibitem{zare1973direct}
RN~Zare, AL~Schmeltekopf, WJ~Harrop, and DL~Albritton.
\newblock A direct approach for the reduction of diatomic spectra to molecular
  constants for the construction of rkr potentials.
\newblock {\em Journal of Molecular Spectroscopy}, 46(1):37--66, 1973.

\bibitem{tully1974collisions}
John~C Tully.
\newblock Collisions of f (2 p 1/2) with h2.
\newblock {\em The Journal of Chemical Physics}, 60(8):3042--3050, 1974.

\bibitem{herschbach1977molecular}
DR~Herschbach.
\newblock Molecular dynamics of chemical reactions.
\newblock In {\em 25th International Congress of Pure and Applied Chemistry},
  pages 61--73. Elsevier, 1977.

\bibitem{herschbach1987closing}
Dudley~R Herschbach.
\newblock Closing remarks. new dimensions in reaction dynamics and electronic
  structure.
\newblock {\em Faraday Discussions of the Chemical Society}, 84:465--478, 1987.

\bibitem{levine2009molecular}
Raphael~D Levine.
\newblock {\em Molecular reaction dynamics}.
\newblock Cambridge University Press, 2009.

\bibitem{schatz1973large}
George~C Schatz, Joel~M Bowman, and Aron Kuppermann.
\newblock Large quantum effects in the collinear f+ h 2→ fh+ h reaction.
\newblock {\em The Journal of Chemical Physics}, 58(9):4023--4025, 1973.

\bibitem{wu1973quantum}
Shiou-Fu Wu, BR~Johnson, and RD~Levine.
\newblock Quantum mechanical computational studies of chemical reactions: Iii.
  collinear a+ bc reaction with some model potential energy surfaces.
\newblock {\em Molecular Physics}, 25(4):839--856, 1973.

\bibitem{truhlar1984resonances}
Donald~G Truhlar.
\newblock Resonances in electron-molecule scattering, van der waals complexes,
  and reactive chemical dynamics.
\newblock 1984.

\bibitem{fernandez2002scattering}
F{\'e}lix Fern{\'a}ndez-Alonso and Richard~N Zare.
\newblock Scattering resonances in the simplest chemical reaction.
\newblock {\em Annual review of physical chemistry}, 53(1):67--99, 2002.

\bibitem{polanyi1995direct}
John~C Polanyi and Ahmed~H Zewail.
\newblock Direct observation of the transition state.
\newblock {\em Accounts of Chemical Research}, 28(3):119--132, 1995.

\bibitem{skodje2000resonance}
Rex~T Skodje, Dimitris Skouteris, David~E Manolopoulos, Shih-Huang Lee, Feng
  Dong, and Kopin Liu.
\newblock Resonance-mediated chemical reaction: F+ h d→ h f+ d.
\newblock {\em Physical Review Letters}, 85(6):1206, 2000.

\bibitem{neumark1984experimental}
DM~Neumark, AM~Wodtke, GN~Robinson, CC~Hayden, and Yuan-Tseh Lee.
\newblock Experimental investigation of resonances in reactive scattering: The
  f+ h 2 reaction.
\newblock {\em Physical review letters}, 53(3):226, 1984.

\bibitem{aoiz1994classical}
F~Javier Aoiz, Luis Ba{\~n}ares, V{\'\i}ctor~J Herrero, V~S{\'a}ez R{\'a}banos,
  K~Stark, and H-J Werner.
\newblock Classical dynamics for the f+ h2→ hf+ h reaction on a new ab initio
  potential energy surface. a direct comparison with experiment.
\newblock {\em Chemical physics letters}, 223(3):215--226, 1994.

\bibitem{castillo1996quantum}
Jesus~F Castillo, David~E Manolopoulos, Klaus Stark, and Hans-Joachim Werner.
\newblock Quantum mechanical angular distributions for the f+ h2 reaction.
\newblock {\em The Journal of chemical physics}, 104(17):6531--6546, 1996.

\bibitem{manolopoulos1993transition}
David~E Manolopoulos, Klaus Stark, Hans-Joachim Werner, Don~W Arnold, Stephen~E
  Bradforth, and Daniel~M Neumark.
\newblock The transition state of the f+ h2 reaction.
\newblock {\em Science}, 262(5141):1852--1855, 1993.

\bibitem{russell1996observe}
Caroline~L Russell and David~E Manolopoulos.
\newblock How to observe the elusive resonances in f+ h2 reactive scattering.
\newblock {\em Chemical physics letters}, 256(4-5):465--473, 1996.

\bibitem{kim2015spectroscopic}
Jongjin~B Kim, Marissa~L Weichman, Tobias~F Sjolander, Daniel~M Neumark, Jacek
  K{\l}os, Millard~H Alexander, and David~E Manolopoulos.
\newblock Spectroscopic observation of resonances in the f+ h2 reaction.
\newblock {\em Science}, 349(6247):510--513, 2015.

\bibitem{yang2019enhanced}
Tiangang Yang, Long Huang, Chunlei Xiao, Jun Chen, Tao Wang, Dongxu Dai,
  Francois Lique, Millard~H Alexander, Zhigang Sun, Dong~H Zhang, et~al.
\newblock Enhanced reactivity of fluorine with para-hydrogen in cold
  interstellar clouds by resonance-induced quantum tunnelling.
\newblock {\em Nature chemistry}, 11(8):744--749, 2019.

\bibitem{qiu2006observation}
Minghui Qiu, Zefeng Ren, Li~Che, Dongxu Dai, Steve~A Harich, Xiuyan Wang,
  Xueming Yang, Chuanxiu Xu, Daiqian Xie, Magnus Gustafsson, et~al.
\newblock Observation of feshbach resonances in the f+ h2→ hf+ h reaction.
\newblock {\em Science}, 311(5766):1440--1443, 2006.

\bibitem{xu2006global}
Chuan-xiu Xu, Dai-qian Xie, and Dong-hui Zhang.
\newblock A global ab initio potential energy surface for f+ h2→ hf+ h.
\newblock {\em Chinese Journal of Chemical Physics}, 19(2):96, 2006.

\bibitem{ren2008probing}
Zefeng Ren, Li~Che, Minghui Qiu, Xingan Wang, Wenrui Dong, Dongxu Dai, Xiuyan
  Wang, Xueming Yang, Zhigang Sun, Bina Fu, et~al.
\newblock Probing the resonance potential in the f atom reaction with hydrogen
  deuteride with spectroscopic accuracy.
\newblock {\em Proceedings of the National Academy of Sciences},
  105(35):12662--12666, 2008.

\bibitem{chen2021quantum}
Wentao Chen, Ransheng Wang, Daofu Yuan, Hailin Zhao, Chang Luo, Yuxin Tan,
  Shihao Li, Dong~H Zhang, Xingan Wang, Zhigang Sun, et~al.
\newblock Quantum interference between spin-orbit split partial waves in the f+
  hd→ hf+ d reaction.
\newblock {\em Science}, 371(6532):936--940, 2021.

\bibitem{zeman2004coherent}
Vlado Zeman, Moshe Shapiro, and Paul Brumer.
\newblock Coherent control of resonance-mediated reactions: F+ hd.
\newblock {\em Physical review letters}, 92(13):133204, 2004.

\bibitem{gong2003entanglement}
Jiangbin Gong, Moshe Shapiro, and Paul Brumer.
\newblock Entanglement-assisted coherent control in nonreactive diatom--diatom
  scattering.
\newblock {\em The Journal of chemical physics}, 118(6):2626--2636, 2003.

\bibitem{yuan2018direct}
Daofu Yuan, Shengrui Yu, Wentao Chen, Jiwei Sang, Chang Luo, Tao Wang, Xin Xu,
  Piergiorgio Casavecchia, Xingan Wang, Zhigang Sun, et~al.
\newblock Direct observation of forward-scattering oscillations in the h+ hd→
  h 2+ d reaction.
\newblock {\em Nature chemistry}, 10(6):653--658, 2018.

\bibitem{alexander2000investigation}
Millard~H Alexander, David~E Manolopoulos, and Hans-Joachim Werner.
\newblock An investigation of the f+ h 2 reaction based on a full ab initio
  description of the open-shell character of the f (2 p) atom.
\newblock {\em The Journal of Chemical Physics}, 113(24):11084--11100, 2000.

\bibitem{alexander2002theoretical}
Millard~H Alexander, Gabriella Capecchi, and Hans-Joachim Werner.
\newblock Theoretical study of the validity of the born-oppenheimer
  approximation in the cl+ h2→ hcl+ h reaction.
\newblock {\em Science}, 296(5568):715--718, 2002.

\bibitem{aspect1982experimental}
Alain Aspect, Philippe Grangier, and G{\'e}rard Roger.
\newblock Experimental realization of einstein-podolsky-rosen-bohm
  gedankenexperiment: a new violation of bell's inequalities.
\newblock {\em Physical review letters}, 49(2):91, 1982.

\bibitem{li2019entanglement}
Junxu Li and Sabre Kais.
\newblock Entanglement classifier in chemical reactions.
\newblock {\em Science advances}, 5(8):eaax5283, 2019.

\bibitem{werner1989quantum}
Reinhard~F Werner.
\newblock Quantum states with einstein-podolsky-rosen correlations admitting a
  hidden-variable model.
\newblock {\em Physical Review A}, 40(8):4277, 1989.

\bibitem{lee2004resonance}
Shih-Huang Lee, Feng Dong, and Kopin Liu.
\newblock A resonance-mediated non-adiabatic reaction.
\newblock {\em Faraday discussions}, 127:49--57, 2004.

\end{thebibliography}
\end{document}